\documentstyle[twoside,fleqn,espcrc2]{article}

\newcommand{\AmS}{{\protect\the\textfont2
  A\kern-.1667em\lower.5ex\hbox{M}\kern-.125emS}}

\def\lapp{\mathbin{\raise2pt \hbox{$<$} \hskip-9pt \lower4pt  
\hbox{$\sim$}}} 
\def\gapp{\mathbin{\raise2pt \hbox{$>$} \hskip-9pt \lower4pt  
\hbox{$\sim$}}} 

\title{BeppoSAX observations of low power radio galaxies: possible detection of
obscured nuclei}

\author{E. Trussoni\address{Osservatorio Astronomico di Torino, 
Pino Torinese, Italy}, %
F. Vagnetti\address{Dipartimento di Fisica, Universit\`a di Roma ``Tor 
Vergata", Roma, Italy}, %
S. Massaglia\address{Dipartimento di Fisica Generale, Universit\`a di Torino, 
Torino, Italy}, %
L. Feretti\address{Istituto di Radioastronomia del CNR, Bologna, Italy}, %
P. Parma$^{\rm d}$, %
R. Morganti$^{\rm d}$, %
R. Fanti\address{Dipartimento di Fisica, Universit\`a di Bologna, Bologna, 
Italy}, 
P. Padovani$^{\rm b,}$\thanks{Present address: ESA, STScI, Baltimora, USA}, %
G. Bodo$^{\rm a}$}

\begin{document}

\begin{abstract}
We present the first results of BeppoSAX observations of a small sample of low
brightness FRI radio galaxies. The flux of all the targets is consistent with a
thermal spectrum, as due to the presence of hot intracluster gas or galactic
corona. Moreover in three sources a non thermal absorbed spectrum can be
present in the MECS spectrum at energies $\gapp \; 7 $ keV, while for a fourth
object a high energy flux has been detected in the PDS instrument at energies
$\gapp \; 15$ keV. This component could be related to the inner AGN surrounded
by an obscuring torus. 

\end{abstract}

\maketitle

\section{INTRODUCTION}

In the unified model for AGN, radio galaxies are usually assumed as parent
objects for radio loud quasars and BL Lac objects. This theory predicts that
the Doppler effect in relativistic jets and the presence of an obscuring torus
around the AGN, with different position angles with respect to the
line of sight, can  explain the
various properties of powerful radio galaxies (FR II) and radio loud quasars
[1,2]. The unification of low luminosity radio galaxies (mostly FR I) with BL
Lac objects in principle does not require a thick absorbing torus [3], however
its presence is not ruled out [1]. This possibility
could be supported by the detection in FR I radio sources of an increasing
X-ray flux at energies above $\approx 5 - 10$ keV, where the contribution from
the thermal component is very low and the photoelectric absorption of the torus
becomes negligible. For the evaluation of the orientation of the AGN - torus  
system, a useful parameter is $R$, the ratio between the core and the lobe radio
fluxes at a given frequency. In sources with low values of $R$ we expect the
absorbing torus to be almost edge-on.

In the framework of this problem, we have observed with BeppoSAX a sample of 5
low power radio galaxies extracted from the 2 Jy sample of Morganti
et al. [4]. The targets (see Table 1), selected  with $R \lapp 1$, are all but
one members of clusters. Their morphology is mostly typical as for FR I radio
sources, however a member of the sample (0625-53) belongs to the class of
low-excitation FR II radio galaxies [5]. 

\begin{table*}[hbt]
 \setlength{\tabcolsep}{1.5pc}
 \newlength{\digitwidth} \settowidth{\digitwidth}{\rm 0}
\catcode`?=\active \def?{\kern\digitwidth}
\caption{The Sample}
\label{tab:effluents}
\begin{tabular*}{\textwidth}{@{}l@{\extracolsep{\fill}}rrrrr}
 \hline
Object    &    Alt. name$\;\;\;$    &   z$\;\;\;\;$  &   m$_V$ &  
R$^{\rm a}\;\;$ &   Envir.   \\
\hline

PKS 0305+03    & 3C78, NGC1218 & $0.029$ & $ 13.84$ & $0.12$ & Isolat. \\
PKS 0620-52    &               & $0.051$ & $ 15.50$ & $0.06$ & Cluster \\
PKS 0625-53    &               & $0.054$ & $ 15.54$ & $0.008$ & A 3391 \\
PKS 0625-35    &  OH 342$\;\;\;\;\;$       & $0.055$ & $ 16.50$ & $0.23$ & A 3392 \\
PKS 1648+05    & 3C38, Her A   & $0.154$ & $ 18.50$ & $0.0004$ & Cluster \\

\hline
\multicolumn{5}{@{}p{120mm}}{$^{\rm a}$ At 2.3 GHz }
\end{tabular*}
\end{table*}

\begin{table*}[hbt]
 \setlength{\tabcolsep}{1.5pc}
\catcode`?=\active \def?{\kern\digitwidth}
\caption{The Observations}
\label{tab:effluents}
\begin{tabular*}{\textwidth}{@{}l@{\extracolsep{\fill}}rrrrr}
  \hline
                  & \multicolumn{2}{r}{MECS$^{\rm a}\;\;\;\;\;\;\;\;\;\;\;$} 
                  & \multicolumn{2}{r}{LECS$\;\;\;\;\;\;\;\;\;\;\;\;\;\;$} \\
\cline{3-3} \cline{5-5}
 Object$\;\;\;\;\;\;\;\;\;$   & Date$\;\;\;\;\;\;$   &  Exp.(s) / Rad.$^{\rm b}$
 &  S. counts &  Exp.(s) / Rad.$^{\rm b}$ &  S. counts  \\
\hline

3C78$\;\;\;\;\;\;\;$ & $7/1/97\;\;\;\;$ &  20594 /$\;\;\;4^{\prime}\;\;\;\;$ &  $494 \pm 21$ & 
 8890 / $8^{\prime}\;\;\;\;$ & $ 142 \pm 27 $ \\ 
0620-52$\;\;\;\;\;\;\;$ & $2/12/96\;\;\;\;$ &  13680 / $10^{\prime}\;\;\;\;$ & 
 $698 \pm 41$ & 4458 / $8^{\prime}\;\;\;\;$ & $ 134 \pm 27 $ \\ 
0625-53$\;\;\;\;\;\;\;$ & $2/12/96\;\;\;\;$ &  13482 / $10^{\prime}\;\;\;\;$ & 
 $4988 \pm 81$ & 3588 / $8^{\prime}\;\;\;\;$ & $ 378 \pm 26 $ \\ 
OH 342$\;\;\;\;\;\;\;$ & $5/10/96\;\;\;\;$ &  17528 / $10^{\prime}\;\;\;\;$ & 
 $1928 \pm 53$ & 8211 / $8^{\prime}\;\;\;\;$ & $ 575 \pm 33 $ \\ 
Her A$\;\;\;\;\;\;\;$ & $28/3/97\;\;\;\;$ &  18836 / $10^{\prime}\;\;\;\;$ & 
 $1620\pm 57$ & 9875 / $8^{\prime}\;\;\;\;$ & $ 504 \pm 30$ \\ 
\hline
\multicolumn{5}{@{}p{120mm}}{$^{\rm a}$ Merging the data of the three 
instruments} \\
\multicolumn{5}{@{}p{120mm}}{$^{\rm b}$ Radius of the region from where the photons
were extracted}

\end{tabular*}
\end{table*}
 
\begin{table*}[hbt]
 \setlength{\tabcolsep}{1.5pc}
\catcode`?=\active \def?{\kern\digitwidth}
\caption{Thermal fits from the MECS (1.5 - 10 keV)}
\label{tab:effluents}
\begin{tabular*}{\textwidth}{@{}l@{\extracolsep{\fill}}rrrr}
\hline
 Object$\;\;\;\;\;\;\;\;\;$   &N$^{\rm a}_{H,\, gal}$   &  
kT (keV) &  $\mu^{\rm b}\;\;\;\;$ &  $\chi^2_{red}$ $\;\;\;\;\;\;\;\;\;\;\;$ \\
\hline
3C78$\;\;\;\;\;\;\;$ & 7.3$\;\;\;\;$ &  $2.7 \pm 0.4$$\;$ &  $0.5 \pm 0.4$ 
& 1.09$\;\;\;\;\;\;\;\;\;\;\;\;\;\;\;$ \\
0620-52$\;\;\;\;\;\;\;$ & 5.2$\;\;\;\;$&  $2.0 \pm 0.3$$\;$ & $0.1 \pm 0.1$
 & 1.08$\;\;\;\;\;\;\;\;\;\;\;\;\;\;\;$ \\
0625-53$\;\;\;\;\;\;\;$ & 5.4$\;\;\;\;$ &  $ 5.4 \pm 0.4$$\;$ & $0.3 \pm 0. 1$ 
& 1.02$\;\;\;\;\;\;\;\;\;\;\;\;\;\;\;$ \\ 
OH 342$\;\;\;\;\;\;\;$ & 7.1$\;\;\;\;$ &  $2.7 \pm 0.2$$\;$ & $0.3 \pm 0.2$ & 
1.13$\;\;\;\;\;\;\;\;\;\;\;\;\;\;\;$ \\ 
Her A$\;\;\;\;\;\;\;$ & 6.3$\;\;\;\;$ &  $4.8 \pm 0.7$$\;$ & $0.3 \pm 0.1$ & 
1.01$\;\;\;\;\;\;\;\;\;\;\;\;\;\;\;$ \\ 
\hline
\multicolumn{5}{@{}p{120mm}}{$^{\rm a}$ $\times 10^{20}$ cm$^{-2}$ (fixed)} \\
\multicolumn{5}{@{}p{120mm}}{$^{\rm b}$ Metallicity in units of the standard
cosmic values}

\end{tabular*}
\end{table*}

\begin{table*}[hbt]
 \setlength{\tabcolsep}{1.5pc}
\catcode`?=\active \def?{\kern\digitwidth}
\caption{Thermal + absorbed power law fits$^{\rm a}$ from the MECS (1.5 - 10
keV)} 
\label{tab:effluents}
\begin{tabular*}{\textwidth}{@{}l@{\extracolsep{\fill}}rrrr}
  \hline
 Object$\;\;\;\;\;\;\;\;\;$   &$\;\;\;\;$   kT (keV) & $\;\;\;\;$ $\chi^2$ (red.)
 &$\;\;\;\;\;\;$ L$^{\rm b}_X$ (ther.)  & 
 L$^{\rm b}_X$ (power l.)$\;\;\;\;\;\;\;$ \\
\hline
3C78$\;\;\;\;\;\;\;$ &  $1.4 \pm 0.3$$\;$ & 0.82$\;\;\;\;$ & 
$ 3.1 \times 10^{42}$ & $ 4.2 \times 10^{42}$$\;\;\;\;\;\;\;\;\;\;$  \\
OH 342$\;\;\;\;\;\;\;$ &  $1.7 \pm 0.2$$\;$ & 1.03$\;\;\;\;$ & 
$ 5.3 \times 10^{43}$ & $ 4.1 \times 10^{43}$$\;\;\;\;\;\;\;\;\;\;$  \\ 
Her A$\;\;\;\;\;\;\;$ & $2.0 \pm 0.4$$\;$ & 0.97$\;\;\;\;$ & 
$ 2.8 \times 10^{44}$ & $ 3.8 \times 10^{44}$$\;\;\;\;\;\;\;\;\;\;$ \\ 
\hline
\multicolumn{5}{@{}p{150mm}}{$^{\rm a}$ Fixed parameters: N$_{H,\, gal}$ as 
in Tab. 3, N$_{H,\, abs} = 10^{23}$ cm$^{-2}$, $\mu=0.3$, $\alpha = 2$} \\

\multicolumn{5}{@{}p{120mm}}{$^{\rm b}$ erg s$^{-1}$ (H$_o$ = 50 km s$^{-1}$ 
Mpc$^{-1}$ )}

\end{tabular*}
\end{table*}

\section{THE OBSERVATIONS}

The details of the MECS and LECS observations are reported in Table 2. The data
have been handled with the SAXDAS pipeline, and the source counts have been
extracted by assuming the standard background evaluated by merging different
exposures of blank fields. The PDS data have been reduced with the XAS
software, and a significative flux has been detected only from the source
0625-53, with a count rate $0.4 \pm 0.1$ s$^{-1}$. The spectral analysis has
been performed with the Xanadu package, rebinning the events in order to have
at least 20 counts per energy channel in the MECS, and 10 counts in the PDS.
Concerning the LECS data, the flux for 3C 78 and  0620-52 is too weak for a
useful analysis. In addition in the other sources some further calibration
analysis is still required.

\section{RESULTS}

We discuss separately the spectral properties deduced from the MECS
and PDS observations.

\subsection{Spectral analysis: MECS}

In the MECS all the sources appear extended, with radii $\sim 150 $ kpc for 3C
78, $\sim 700 - 800$ kpc for 0625-53, OH 342 and 0620-52, and $\sim 1.5$ Mpc
for Her A. This is expected taking into account that four targets
belong to clusters, while 3C 78 is an isolated galaxy but embedded in
a hot galactic corona. Consistently, from the spectral analysis we see that all
the sources can be satisfactorily fitted with a thermal spectrum
(Raymond-Smith), where the iron line appears quite evident. The spectral
parameters, with the hydrogen column density kept fixed to its galactic value
N$_{H, \, gal}$, are reported in Table 3. We have also tested that fits 
with power law spectra are much worse.
 
However, even though the thermal spectrum is consistent with
the data, the following points must be remarked:

1) In three sources (3C78, OH 342 and Her A) a count excess is evident
at high energies ($ \gapp \; 7$ keV).

2) The temperature obtained for the halo of  3C 78 is higher than expected
for hot galactic coronae ($\approx 0.5 - 1.5$ keV [6,7,8]).

3) The temperature of the intracluster gas of Her A is higher than obtained 
from a Rosat observation in the soft X-ray energy band [9].

These issues suggest that a second component can be present in the spectrum of
the sources. In the framework of our initial discussion, we have assumed that
this second component may be related to the non-thermal emission from the
central nucleus, surrounded by an obscuring torus seen edge-on. 
 
Due to the low photon flux, the fit with a double spectrum does not provide
useful results unless some parameters are fixed. We have seen that the best
fits are obtained only whether a high absorption is assumed for the non-thermal
component. Therefore, besides the galactic hydrogen column density N$_{H,\,
gal}$, we have fixed the value of the hydrogen column density of the absorbing
torus N$_{H, \, abs} = 10^{23}$ cm$^{-2}$, the metallicity of the thermal
component $\mu =0.3$ (in units of the standard cosmic values), and the photon
index of the power law spectrum $\alpha=2$. For three sources (3C78, OH 342 and
3C78) the presence of an absorbed non-thermal component is consistent with the
data, as we can see in the unfolded spectra plotted in Figs. 1 and 2. The
luminosity associated with the non-thermal emission from these AGN is
$\approx 4 \times 10^{42 - 44}$ erg s$^{-1}$ in the MECS energy range (it is
quite outstanding  the high nuclear luminosity of Her A). With respect to the
case of a single thermal spectrum, now the temperature of the hot gas in these
objects is quite lower (see Table 4). 
\begin{figure}[htb]
\vspace{9pt}
\includegraphics{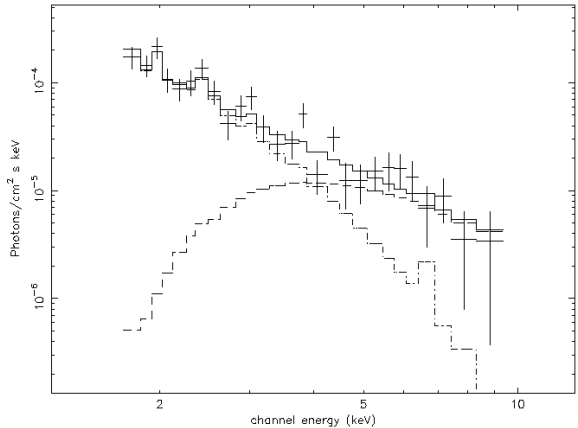}
\vspace{4.3 cm}
\caption{The MECS unfolded, two-component spectrum (thermal + absorbed power
law) of 3C 78. The spectral parameters of the fit are reported in Table 4} 
\end{figure}

\begin{figure}[htb]
\vspace{9pt}
\includegraphics{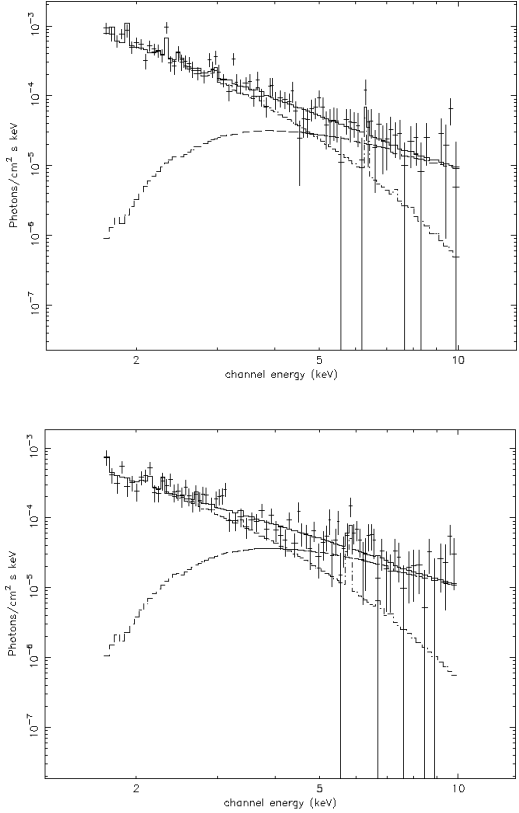}
\vspace{9.9 cm}
\caption{The MECS unfolded, two-component spectrum (thermal + absorbed power
law) of OH 342 (upper panel) and Her A (lower panel). The spectral parameters
of the fit are reported in Table 4} 
\end{figure}

\begin{figure}[htb]
\vspace{9pt}
\includegraphics{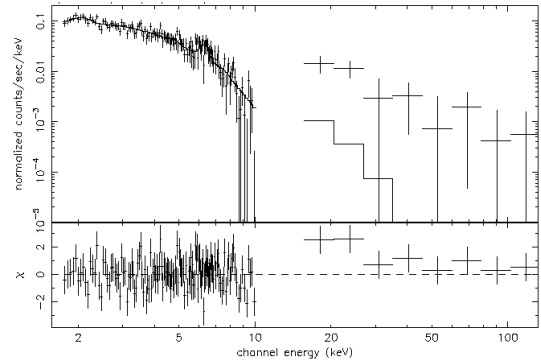}
\vspace{4.3 cm}
\caption{MECS and PDS spectrum of 0625-53. The fit refers to the thermal 
spectrum of the MECS data  (see Table 3)} 
\label{fig:largenenough}
\end{figure}

\subsection{Spectral analysis: PDS}

In the MECS, the data of 0625-53 (associated with the cluster Abell 3391) are
fully dominated by the emission of the intracluster gas. This thermal spectrum
cannot be related to the high energy flux detected in the PDS, as we can see
in Fig. 3. Assuming that this  emission originates from nucleus of the
radio galaxy, and is non-thermal with a photon index $\alpha=2$, we deduce a
luminosity $4.1 \times 10^{44}$ erg s$^{-1}$, in the energy range $15 - 150$
keV. This also implies that this spectral component must have a cut-off at
energies $\sim 10$ keV, consistent with a hydrogen column density N$_{H, \,
abs} \gapp 10^{24}$ cm$^{-2}$. We must remark however that we cannot exclude at
the moment that the detected flux originates from a foreground hard X-ray 
source in the field of view of the PDS (survey still in progress). 

\section{SUMMARY}

From the MECS observations of the five radio galaxies of our sample, an
absorbed non-thermal emission seems to be present in 3C78, OH 342 and Her A,
that can be associated with the active nucleus surrounded by a thick obscuring
torus. A similar interpretation can also hold for the radio galaxy 0625-53, but
at much higher energies. Should this scenario be confirmed by detailed
observations of a larger sample of low power radio galaxies, then the unified
scheme proposed for FR II - radio loud quasars (Doppler beaming plus absorbing
torus) could be also valid for FR I - Bl Lac objects.

\end{document}